\documentclass[11pt,twoside]{article}

\usepackage{asp2014}

\aspSuppressVolSlug
\resetcounters

\begin{document}

\title{Observing, calibrating and reducing near-infrared imaging mosaics}

\author{Kirill Grishin,$^{1,2}$ Igor Chilingarian,$^{3,2}$}
\affil{$^1$Universit\'e de Paris, CNRS, Astroparticule et Cosmologie, Paris, France}
\affil{$^2$Sternberg Astronomical Institute, M.V.~Lomonosov Moscow State University,  Moscow, Russia}
\affil{$^2$Center for Astrophysics - Harvard and Smithsonian, Cambridge, USA}

% This section is for ADS Processing.  There must be one line per author. paperauthor has 9 arguments.
\paperauthor{Kirill Grishin}{grishin@voxastro.org}{0000-0003-3255-7340}{Université de Paris, CNRS, Astroparticule et Cosmologie}{}{Paris}{}{75013}{France}
\paperauthor{Igor Chilingarian}{igor.chilingarian@cfa.harvard.edu}{0000-0002-7924-3253}{Center for Astrophysics - Harvard and Smithsonian}{}{Cambridge}{}{02138}{USA}

\begin{abstract}
In near-infrared bands, co-adding and tiling of astronomical imaging datasets require a sufficiently high calibration quality (flat fielding, background subtraction). Here we present a complete workflow for obtaining imaging mosaics with the MMT and Magellan Infrared Spectrograph (MMIRS) operated at the 6.5-m MMT in Arizona and open-source add-on tools developed for the MMIRS pipeline for preparation and data reduction of mosaic observations. We describe pre-observing actions, such as design of dithering patterns and mosaic layouts and post-processing steps to perform absolute astrometric and photometric calibration, and also generate HiPS maps to display the final data product in Aladin / Aladin Lite.
\end{abstract}

% These lines show examples of subject index entries. At this stage these have to commented
% out, and need to be on separate lines. Eventually, they will be automatically uncommented
% and used to generate entries in the Subject Index at the end of the Proceedings volume.
% Don't leave these in! - replace them with ones relevant to your paper.
%\ssindex{FOOBAR!conference!ADASS 2019}
%\ssindex{FOOBAR!organisations!ASP}

% These lines show examples of ASCL index entries. At this stage these have to commented
% out, and need to be on separate lines. Eventually, they will be automatically uncommented
% and used to generate entries in the ASCL Index at the end of the Proceedings volume.
% The ascl.py command will scan your paper on possible code names.
% Don't leave these in! - replace them with ones relevant to your paper.
%\ooindex{FOOBAR, ascl:1101.010}

\section{Introduction}
Co-adding and tiling of astronomical imaging datasets allow one to investigate low-surface brightness features of extended objects such as galaxies, nebulae, comets etc. However, near-infrared CMOS detectors compared to optical CCDs contain a lot of imperfections: pixel-to-pixels sensitivity variations are high (30\%) and wavelength-dependent; there is persistence from bright sources, cross-talk from different read-out channels. To overcome these difficulties, while taking a series of NIR images the telescope should slightly change it position in accordance with the chosen dithering pattern, which in many cases is a random sequence. Another difficulty comes from the night sky background level that is much higher compared to optical wavelengths and at the same time its spectral content changes spatially and temporally.

Here we present an infrared imaging survey conducted with the MMIRS instrument at the 6.5-m. MMT telescope \citep{2012PASP..124.1318M} for the Coma cluster that follows up our optical spectroscopic observational campaign carried out with the Binospec multi-object spectrograph \citep{2019PASP..131g5004F} to study the evolution and properties of dwarf galaxies in clusters. These data are used for SED modelling of ultra-diffuse galaxies and accurate determination of their structural parameters and stellar masses \citep{2021NatAs.tmp..208G, 2019arXiv190913460G, 2019ApJ...884...79C}.

\section{Observing strategy}
Typically, an imaging mosaic is taken as a set of separate images that have their own dithering patterns for each field. However, these fields are usually much closer to each other than the maximal offset size allowed by the telescope without re-acquiring the target (1~deg for the MMT). Therefore, instead of putting separate observation for each field in the telescope scheduling system, one can make one observation for several adjacent fields with a custom dithering pattern for it, which will include dithering patterns for each single field concatenated into one large single pattern. Changing the field rotator angle at a large telescope is usually much more time-consuming than short slews, therefore it is recommended to limit the number of position angle values to 2--3.  

For the Coma cluster survey we split 36 fields covering the footprint of 8 Binospec slit masks into two groups with PA values of 0$^{\circ}$ (11 fields) and 340$^{\circ}$ (25 fields). 4 fields at PA=340$^{\circ}$ fill into the footprint of archival imaging datasets obtained in $J$ and $K$ bands with WIRCam at CFHT, so they were excluded from the dithering patterns for these bands and retained only in the $H$ band. All fields at PA=0$^{\circ}$ were joined into one dithering pattern since they are spanning an area of the sky 14~arcmin across. Fields at PA=340$^{\circ}$ cover a larger area containing the footprint of two Binospec masks (``Coma 1'' and ``Coma B'') located far from the cluster center. To prevent the telescope from getting close to the dithering pattern size limit these fields were split into two patterns. 

\section{Data reduction}
The data reduction was done using the standard MMIRS pipeline \citep{2015PASP..127..406C} in the imaging mode. For each field during the reduction process the images from the other fields obtained in the same night containing no extended objects (e.g. larger galaxies) were used as offset sky images from flat field construction. 

The large number of fields makes the manual preparation of pipeline configurations exhausting and increases the time investments into whole data reduction process. To automate this process we have developed a {\sc Python}-based wrapper that splits the list of raw pre-processed images obtained in a each dithering pattern into a set of images that belong to a single field. For the stack co-adding, MMIRS pipeline estimates the centroid position of a reference star whose coordinates on the first image in the sequence are set manually by the user. This part of configuration preparation process was also automatised in the wrapper by applying {\sc SExtractor} package on raw images and filtering its output source catalogs.

As an output pipeline provides flatfielded co-added FITS images with World Coordinate System (WCS) that came from the Telescope Control System (TCS).

\section{Sky background correction}
The standard data reduction includes a background subtraction where the background shape is considered flat. However, in NIR bands, especially $H$, the air glow has variations on a small spatial scales (several arcmin) causing the gradient of the sky background across the MMIRS 7'$\times$7' field of view. Given that the air grow OH emission lines may change on short ($<$ 1~min) timescales, the co-adding of even a substantial (10$\dots$20) number of frames does not remove the sky gradient.

For sky background correction procedure on each frame in a offset-corrected stack we mask objects using the co-added image returned by the pipeline. Unmasked pixels are fitted with a low-order 2D polynomial, which is then subtracted from the frame. 

\begin{figure}
\vskip -0.45cm
\centering
\begin{minipage}[b]{0.55\linewidth}
\includegraphics[width=1.00\hsize]{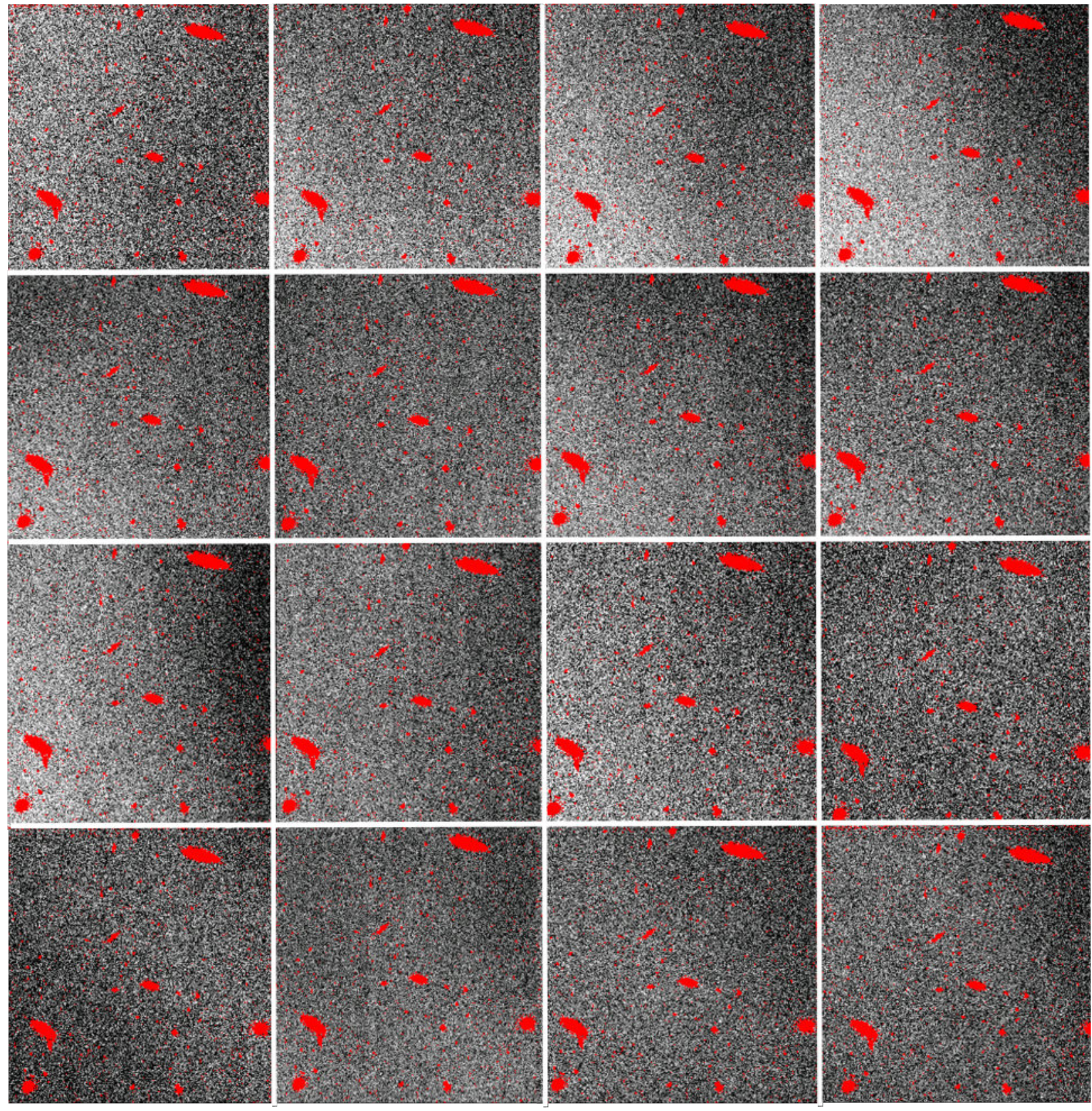}
\end{minipage}
\begin{minipage}[b]{0.29\linewidth}
\includegraphics[width=1.00\hsize]{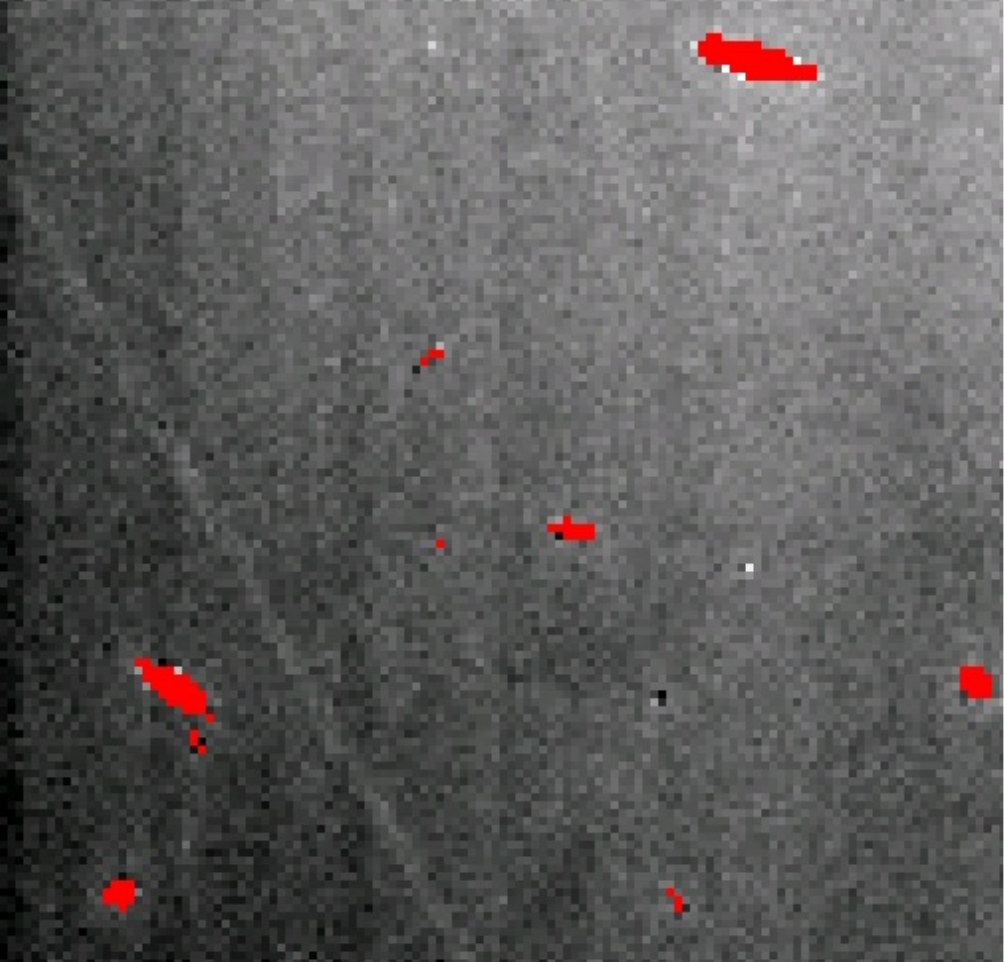}\\
\includegraphics[width=1.00\hsize]{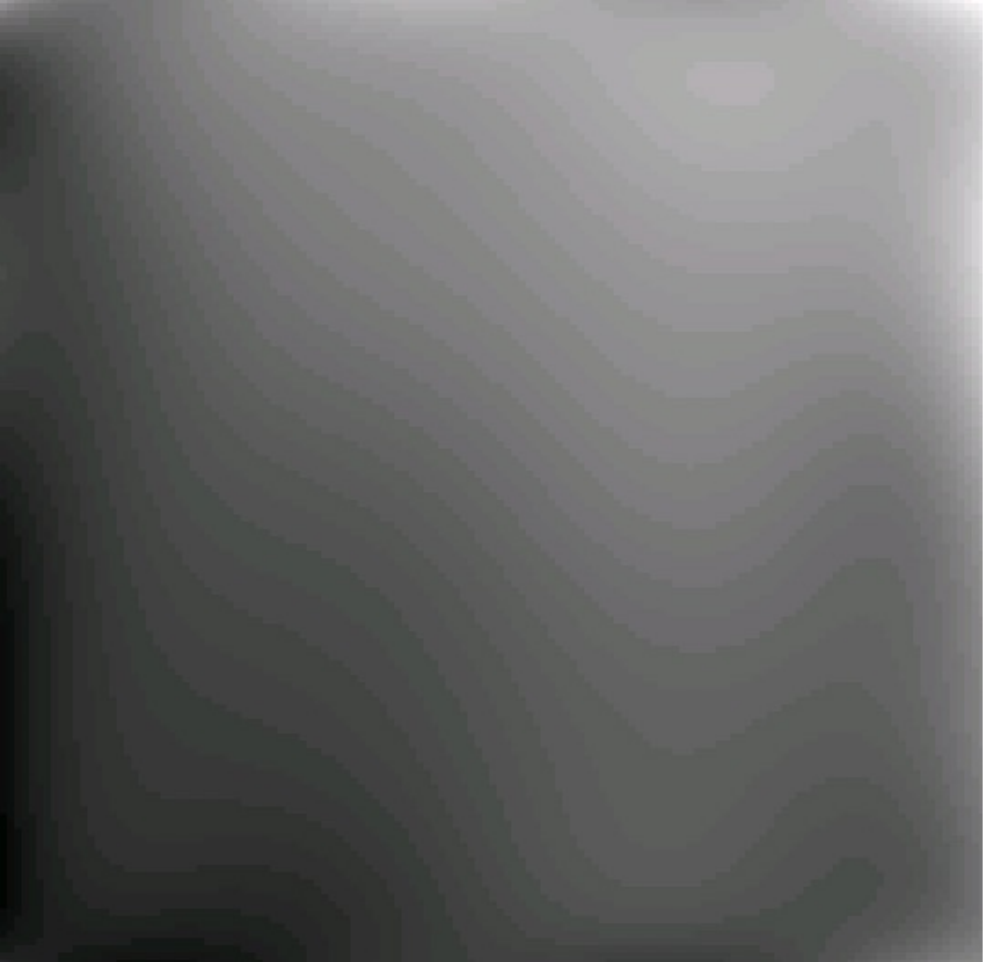}
\end{minipage}
\vskip -0.3cm
\caption{Left: Sky background patterns (masked objects are shown in red) of a sequence of 16 images in the $H$ band. The exposure time for each image is 15 sec. Right: an example of a binned background pattern (upper frame) approximated by a low-order 2D polynomial (lower frame).
\label{bgr_var}}
\vskip -0.70cm
\end{figure}

\section{Astrometric and photometric calibration}
Reduced images returned by the MMIRS pipeline contain the WCS originating the TCS which may be offset by as much as 15'', so at this stage the fields cannot be tiled and the WCS correction is required. We developed a dedicated {\sc Python} package that retrieves a subset from a reference astrometric catalog selected by a user from the following options: GAIA, UKIDSS \citep{2007MNRAS.379.1599L}, PanSTARRS and 2MASS \citep{2006AJ....131.1163S}. For the retrieval procedure it utilizes the {\sc PyVO} methods. Then, the {\sc SExtarctor} \citep{1996A&AS..117..393B} is applied to the image to create a source catalog which is then cross-matched with the reference catalog. The pairs, containing celestial and image coordinates are used for WCS parameters determination using {\sc Lmfit} package \citep{2014zndo.....11813N}. This procedure is repeated with outlier rejection on each iteration until some given threshold on $\chi^2$ is reached.

For the photometric zero-point calibration our package also retrieves a subset of a reference catalog and performs the cross-match with the source catalog for the image. Then a value of the photometric zero-point is determined using a linear regression.

\section{Image tiling using Hierarchical Progressive Surveys (HiPS)}
\begin{figure}
\vskip -0.45cm
\centering
\begin{minipage}[b]{0.61\linewidth}
\includegraphics[width=\hsize]{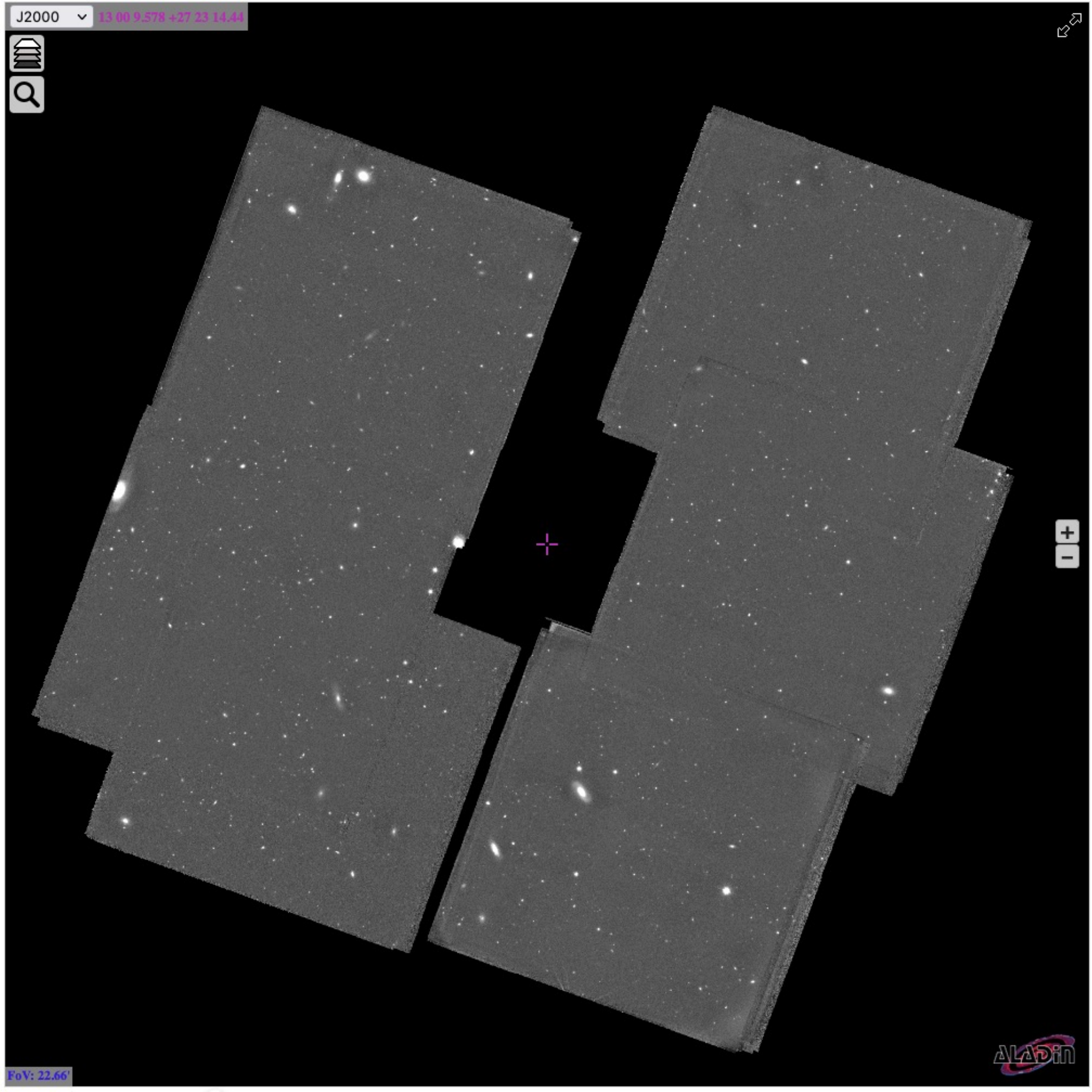}
\end{minipage}
\begin{minipage}[b]{0.37\linewidth}
\includegraphics[width=\hsize]{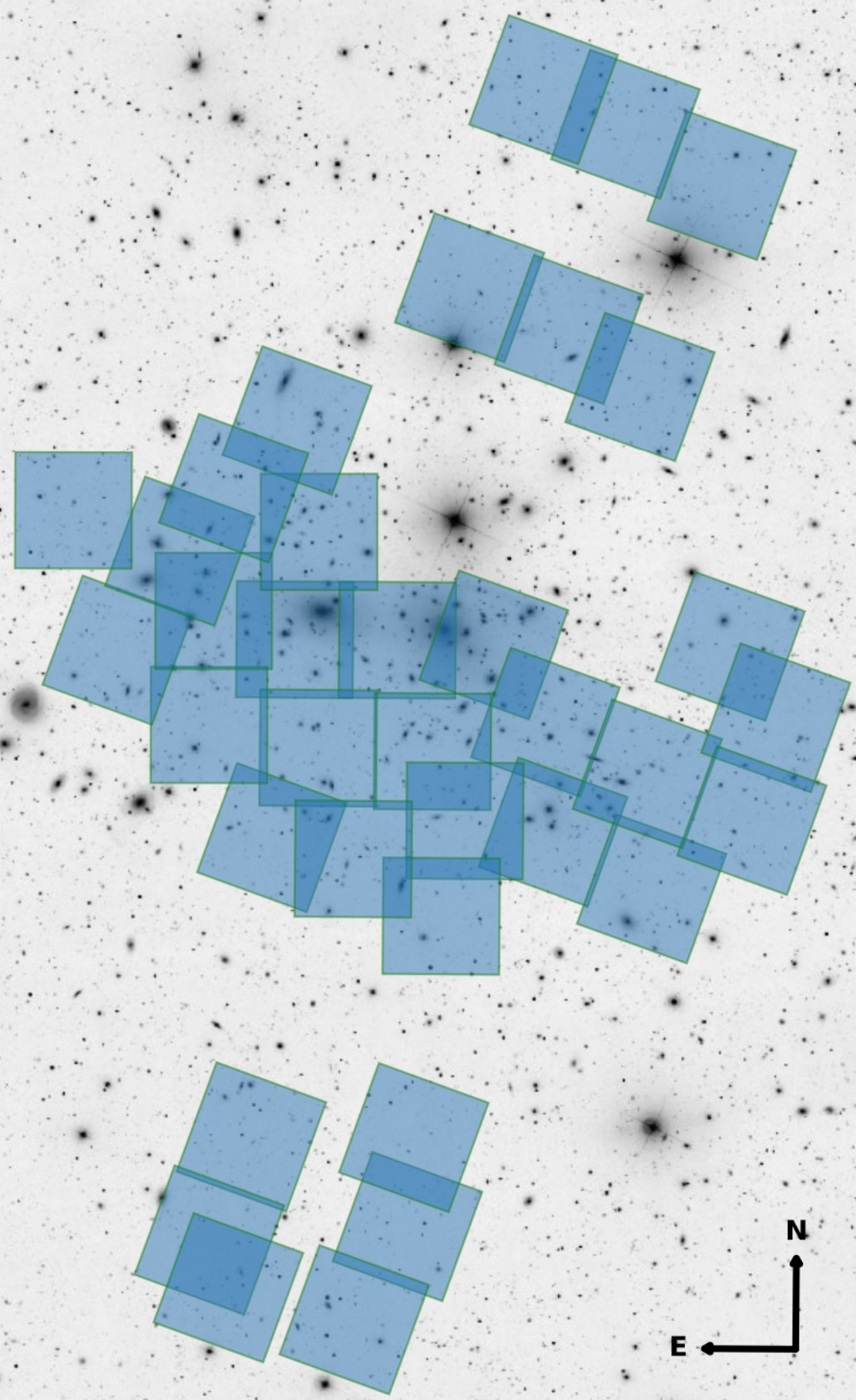}
\end{minipage}

\vskip -0.3cm
\caption{Left: A mosaic of 6 MMIRS Coma cluster fields in the Aladin Lite interface. Right: Positions of all observed fields on top of a Coma cluster image.
\label{aladin_lite}}
\vskip -0.45cm
\end{figure}

For visualization and basic data analysis purposes, tiling of all 36 fields into one {\sc FITS} file may not be the best solution because its size will be very large and the manipulations with such files is computationally and network-heavy. Hence, we use HiPS \citep{2015A&A...578A.114F}, a multi-order HEALPix tessellation for astronomical imaging surveys, which can be easily viewed using Aladin desktop \citep{2000A&AS..143...33B} and embedded into any web-page (Fig.~\ref{aladin_lite}) using Aladin Lite \citep{2014ASPC..485..277B}. This visualization technique broadens the opportunities of this imaging survey to be used as a value-added dataset in many other projects, including RCSED \citep{RCSED}. 

\acknowledgements This project is supported by the RScF Grant 19-12-00281 and the Interdisciplinary Scientific and Educational School of Moscow University ``Fundamental and Applied Space Research''. KG is grateful to the ADASS-XXXI organizing committee for providing financial aid to support his attendance of the conference.

\bibliography{O6-001_arxiv}

\begin{thebibliography}{}
\expandafter\ifx\csname natexlab\endcsname\relax\def\natexlab#1{#1}\fi
\expandafter\ifx\csname url\endcsname\relax
  \def\url#1{\texttt{#1}}\fi
\expandafter\ifx\csname urlprefix\endcsname\relax\def\urlprefix{URL }\fi
\providecommand{\eprint}[2][]{\url{#2}}

\bibitem[{{Bertin} \& {Arnouts}(1996)}]{1996A&AS..117..393B}
{Bertin}, E., \& {Arnouts}, S. 1996, \aaps, 117, 393

\bibitem[{{Boch} \& {Fernique}(2014)}]{2014ASPC..485..277B}
{Boch}, T., \& {Fernique}, P. 2014, in Astronomical Data Analysis Software and
  Systems XXIII, edited by N.~{Manset}, \& P.~{Forshay}, vol. 485 of
  Astronomical Society of the Pacific Conference Series, 277

\bibitem[{{Bonnarel} et~al.(2000){Bonnarel}, {Fernique}, {Bienaym{\'e}},
  {Egret}, {Genova}, {Louys}, {Ochsenbein}, {Wenger}, \&
  {Bartlett}}]{2000A&AS..143...33B}
{Bonnarel}, F., {et~al.} 2000, \aaps, 143, 33

\bibitem[{{Chilingarian} et~al.(2015){Chilingarian}, {Beletsky}, {Moran},
  {Brown}, {McLeod}, \& {Fabricant}}]{2015PASP..127..406C}
{Chilingarian}, I., {et~al.} 2015, \pasp, 127, 406. \eprint{1503.07504}

\bibitem[{{Chilingarian} et~al.(2019){Chilingarian}, {Afanasiev}, {Grishin},
  {Fabricant}, \& {Moran}}]{2019ApJ...884...79C}
{Chilingarian}, I.~V., {et~al.} 2019, \apj, 884, 79. \eprint{1901.05489}

\bibitem[{{Chilingarian} et~al.(2017){Chilingarian}, {Zolotukhin}, {Katkov},
  {Melchior}, {Rubtsov}, \& {Grishin}}]{RCSED}
--- 2017, \apjs, 228, 14. \eprint{1612.02047}

\bibitem[{{Fabricant} et~al.(2019){Fabricant}, {Fata}, {Epps}, {Gauron},
  {Mueller}, {Zajac}, {Amato}, {Barberis}, {Bergner}, {Brennan}, {Brown},
  {Chilingarian}, {Geary}, {Kradinov}, {McLeod}, {Smith}, \&
  {Woods}}]{2019PASP..131g5004F}
{Fabricant}, D., {et~al.} 2019, \pasp, 131, 075004. \eprint{1905.03320}

\bibitem[{{Fernique} et~al.(2015){Fernique}, {Allen}, {Boch}, {Oberto},
  {Pineau}, {Durand}, {Bot}, {Cambr{\'e}sy}, {Derriere}, {Genova}, \&
  {Bonnarel}}]{2015A&A...578A.114F}
{Fernique}, P., {et~al.} 2015, \aap, 578, A114. \eprint{1505.02291}

\bibitem[{{Grishin} et~al.(2021){Grishin}, {Chilingarian}, {Afanasiev},
  {Fabricant}, {Katkov}, {Moran}, \& {Yagi}}]{2021NatAs.tmp..208G}
{Grishin}, K.~A., {et~al.} 2021, Nature Astronomy. \eprint{2111.01140}

\bibitem[{{Grishin} et~al.(2019){Grishin}, {Chilingarian}, {Afanasiev}, \&
  {Katkov}}]{2019arXiv190913460G}
--- 2019, arXiv e-prints, arXiv:1909.13460. \eprint{1909.13460}

\bibitem[{{Lawrence} et~al.(2007){Lawrence}, {Warren}, {Almaini}, {Edge},
  {Hambly}, {Jameson}, {Lucas}, {Casali}, {Adamson}, {Dye}, {Emerson},
  {Foucaud}, {Hewett}, {Hirst}, {Hodgkin}, {Irwin}, {Lodieu}, {McMahon},
  {Simpson}, {Smail}, {Mortlock}, \& {Folger}}]{2007MNRAS.379.1599L}
{Lawrence}, A., {et~al.} 2007, \mnras, 379, 1599. \eprint{astro-ph/0604426}

\bibitem[{{McLeod} et~al.(2012){McLeod}, {Fabricant}, {Nystrom}, {McCracken},
  {Amato}, {Bergner}, {Brown}, {Burke}, {Chilingarian}, {Conroy}, {Curley},
  {Furesz}, {Geary}, {Hertz}, {Holwell}, {Matthews}, {Norton}, {Park}, {Roll},
  {Zajac}, {Epps}, \& {Martini}}]{2012PASP..124.1318M}
{McLeod}, B., {et~al.} 2012, \pasp, 124, 1318. \eprint{1211.6174}

\bibitem[{{Newville} et~al.(2014){Newville}, {Stensitzki}, {Allen}, \&
  {Ingargiola}}]{2014zndo.....11813N}
{Newville}, M., {et~al.} 2014, {LMFIT: Non-Linear Least-Square Minimization and
  Curve-Fitting for Python}

\bibitem[{{Skrutskie} et~al.(2006){Skrutskie}, {Cutri}, {Stiening}, {Weinberg},
  {Schneider}, {Carpenter}, {Beichman}, {Capps}, {Chester}, {Elias}, {Huchra},
  {Liebert}, {Lonsdale}, {Monet}, {Price}, {Seitzer}, {Jarrett}, {Kirkpatrick},
  {Gizis}, {Howard}, {Evans}, {Fowler}, {Fullmer}, {Hurt}, {Light}, {Kopan},
  {Marsh}, {McCallon}, {Tam}, {Van Dyk}, \& {Wheelock}}]{2006AJ....131.1163S}
{Skrutskie}, M.~F., {et~al.} 2006, \aj, 131, 1163

\end{thebibliography}

\end{document}